\documentclass[a4paper,11pt]{article}
\usepackage{pos}

\usepackage{multicol}
\usepackage{multirow}
\usepackage{graphicx,subfigure}
\usepackage{lineno}     
\modulolinenumbers[2]
\switchlinenumbers   

\title{Testing CPT symmetry via precise mass measurements of multi-strange baryons in ALICE}

\author*[a]{Romain Schotter}

 \onbehalf{for the ALICE collaboration}

\affiliation[a]{Stefan Meyer Institute for Subatomic Physics, Austrian Academy of Sciences,\\
  Dominikanerbastei 16, Vienna, Austria}

\emailAdd{romain.schotter@cern.ch}

\abstract{These proceedings present measurements of the $\Xi^{-}$, $\overline{\Xi}^{+}$, $\Omega^{-}$, $\overline{\Omega}^{+}$ masses and mass differences between particle and anti-particle, in pp collisions at $\sqrt{s} = 13$ TeV collected by the ALICE Collaboration during the LHC Run 2. Relying on a data sample much larger than those used in past experiments, the results improve significantly the values obtained from previous measurements and thus offer the opportunity to test directly the CPT symmetry to an unprecedented level of precision in the multi-strange baryon sector.}

\FullConference{%
42nd International Conference on High Energy Physics (ICHEP2024)\\
18-24 July 2024\\
Prague, Czech Republic\\
}

\begin{document}
\maketitle

\section{Introduction}
\label{intro}

In our present understanding, the Lorentz and CPT invariances certainly stand as the most fundamental symmetries in Physics. While the former states that the laws of Physics are the same in all inertial frames, the latter imposes that the laws of Physics remain unchanged under the simultaneous transformations of charge conjugation (C), space inversion (P) and time reversal (T). These symmetries are closely connected by the so-called CPT theorem which states that  that any unitary, local, Lorentz-invariant quantum field theory in a flat Minkowski spacetime must also be CPT invariant \cite{kosteleckyStatusCPT1998}. As a consequence, a violation of the CPT symmetry is generally considered to imply the breaking of the Lorentz symmetry (and vice versa) \cite{sozziTestsDiscreteSymmetries2019}, although one could be violated while preserving the other by dropping some of the other assumptions of the CPT theorem namely the energy positivity, local interactions, finite
spin, etc. Furthermore, due to the charge conjugation, the CPT symmetry also draws a link between particles and anti-particles, imposing that they should have the same properties such as invariant mass, lifetime, etc \cite{lehnertCPTSymmetryIts2016}.

The CPT symmetry has been extensively tested to a high degree of precision in a large variety of experiments. While the sharpest figure of merit is the measure of the relative mass difference between $\text{K}^{0}$ and $\overline{\text{K}}^{0}$ with a precision of about $ 10^{-19}$, the extension to the multi-strange baryon domain still lacks of precise measurements. The only measurements of this nature quoted by the Particle Data Group (PDG) \cite{particledatagroupReviewParticlePhysics2022} dates back to 2006 at LEP-1 \cite{abdallahMassesLifetimesProduction2006}, and to 1998 at the E571 spectrometer~\cite{chanMeasurementPropertiesOverline1998}; both rely on limited statistics: approximately 2500 (2300) $\Xi^{-}$ ($\overline{\Xi}^{+}$) and 6323 (2607) $\Omega^{-}$ ($\overline{\Omega}^{+}$) respectively, as shown in Tab.~\ref{tab:PDGcharac}.

The hadron masses also play a crucial role in non-perturbative QCD, and most particularly, in lattice QCD (lQCD) calculations. They are used as anchor points for determining the lattice spacing, thus allowing to convert predictions in dimensionless lattice units into dimensionful physical units. Several considerations guide the choice of the particles of interest; amongst these, on the one hand, the higher the strangeness content, the more precise the physical scale determination. The $\Omega$ baryon thus stands as a prime candidate although, on the other hand, its mass value is less precise than the $\Xi$ baryon one \cite{durrInitioDeterminationLight2008}. Indeed, while the $\Xi$ mass measurement has been updated at the same time as the mass difference value in 2006, the most recent measurement of the $\Omega$ mass has been performed almost 40 years ago at Fermilab \cite{hartouniNclusiveRoductionEnsuremath1985} using only 100 (72) $\Omega^{-}$ ($\Omega^{+}$).

In these proceedings, we present measurements of the mass and mass difference between particle and anti-particle of the $\Xi^{-}$ and $\overline{\Xi}^{+}$, and $\Omega^{-}$ and $\overline{\Omega}^{+}$ baryons. The data samples are much larger than those exploited previously: $\sim$ 2 400 000 $(\Xi^{-} + \overline{\Xi}^{+})$ and $\sim$ 130 000  $(\Omega^{-} + \overline{\Omega}^{+})$, with a small background. Beyond significantly improving the absolute mass values, these measurements also provide a direct test of the CPT symmetry to an unprecedented level of precision in the multi-strange baryon sector.

\begin{table}[h]
    \centering
	\caption{Particle properties as of 2022, listed into~\cite{particledatagroupReviewParticlePhysics2022}. Here, the mass difference refers to the normalized one, namely $(M_{\overline{\text{part}}} - M_{\text{part}}) / M_{\text{average}}$.}\label{tab:PDGcharac}
    \footnotesize
    \begin{tabular}{cccc|cc}
    \noalign{\smallskip}\hline \hline \noalign{\smallskip}
    \multirow{3}{*}{Particle} & \multirow{3}*{Quark content} & \multirow{3}*{Last mass measurement (MeV$/c^{2}$)} & \multirow{3}*{Sample} & Last mass & \multirow{3}*{Sample}\\
    & & & & difference measurement & \\
    & & & & ($\times 10^{-5}$) & \\
    \noalign{\smallskip}\hline \noalign{\smallskip}
    $\Xi^{-}$ & $dss$ & 1321.70 $\pm$ (stat.)0.08 $\pm$ (syst.)0.05 & 2500 & \multirow{2}*{2.5 $\pm$ (tot.)8.7} & 2500\\
	$\overline{\Xi}^{+}$ & $\bar{d}\bar{s}\bar{s}$ & 1321.73 $\pm$ (stat.)0.08 $\pm$ (syst.)0.05 & 2300 & & 2300\\
    \noalign{\smallskip}\hline \noalign{\smallskip}
    $\Omega^{-}$ & $sss$ & 1672 $\pm$ (tot.)1 & 100 & \multirow{2}*{1.44 $\pm$ (tot.)7.98} & 6323\\ 
    $\overline{\Omega}^{+}$ & $\bar{s}\bar{s}\bar{s}$ & 1673 $\pm$ (tot.)1 & 72 & & 2607 \\ 
	\noalign{\smallskip}\hline \hline \noalign{\smallskip}
    \end{tabular}
\end{table}


\section{Detector setup and data sample}
\label{sec-1}

The measurement is performed using the central detectors of ALICE \cite{alicecollaborationALICEExperimentCERN2008} at the LHC. The Inner Tracking System (ITS) --- composed of six layers of silicon pixel, drift and strip detectors during the LHC Runs 1 and 2 --- allows to reconstruct the primary and secondary vertices with a spatial resolution of 50 $\mu$m. The Time Projection Chamber (TPC) is the main tracking device, offering a momentum resolution of 1 \%, as well as a robust particle identification of pions, kaons and protons based on their energy loss in the detector. Both detectors are embedded inside the L3 magnet, a large solenoid magnet providing different magnetic field configurations ($B=$+0.5, -0.5, -0.2 T).

In this analysis, all the pp collisions at a centre-of-mass energy $\sqrt{s}$ = 13 TeV, collected in 2016, 2017 and 2018 at the nominal magnetic field value ($B= \pm$0.5 T), are exploited. This represents about $2.2 \times 10^{9}$ minimum-bias events.

\section{Data analysis}
\label{sec-2}

The charged $\Xi$ and $\Omega$ baryons are studied using the ALICE detectors at mid-rapidity ($\left| y \right| < 0.5$) in their cascade decay channel: $\Xi^{\pm} \rightarrow \pi^{\pm} \Lambda \rightarrow \pi^{\pm} \pi^{\pm} p^{\mp}$ (with a branching ratio BR = 63.9 \%) and $\Omega^{\pm} \rightarrow {\rm K}^{\pm} \Lambda \rightarrow {\rm K}^{\pm} \pi^{\pm} p^{\mp}$ (BR = 43.4 \%). These decays are reconstructed through topological reconstruction: oppositely charged secondary tracks are paired in order to form a $\Lambda$ candidate, which is then matched to a secondary pion or kaon track. To reduce the combinatorial background, various geometric and kinematic selections are exploited (similarly to those in \cite{alicecollaborationMultiplicityDependenceMulti2020}).

For each candidate, the invariant mass is computed under the $\Xi$ and $\Omega$ hypotheses. The mass of the multi-strange baryons are then extracted via a fit of the corresponding invariant mass distribution; in the standard approach, the peak is modelled by a triple Gaussian, and the background by an exponential. The measured mass is given by the position of the mean ($\mu$) of the triple Gaussian function, and the width ($\sigma$) provides an estimation of the mass resolution. The statistical uncertainties on both quantities correspond to the errors returned by the fit procedure.

\begin{figure}[t]
\subfigure[]{
	\includegraphics[width=7.75cm,clip]{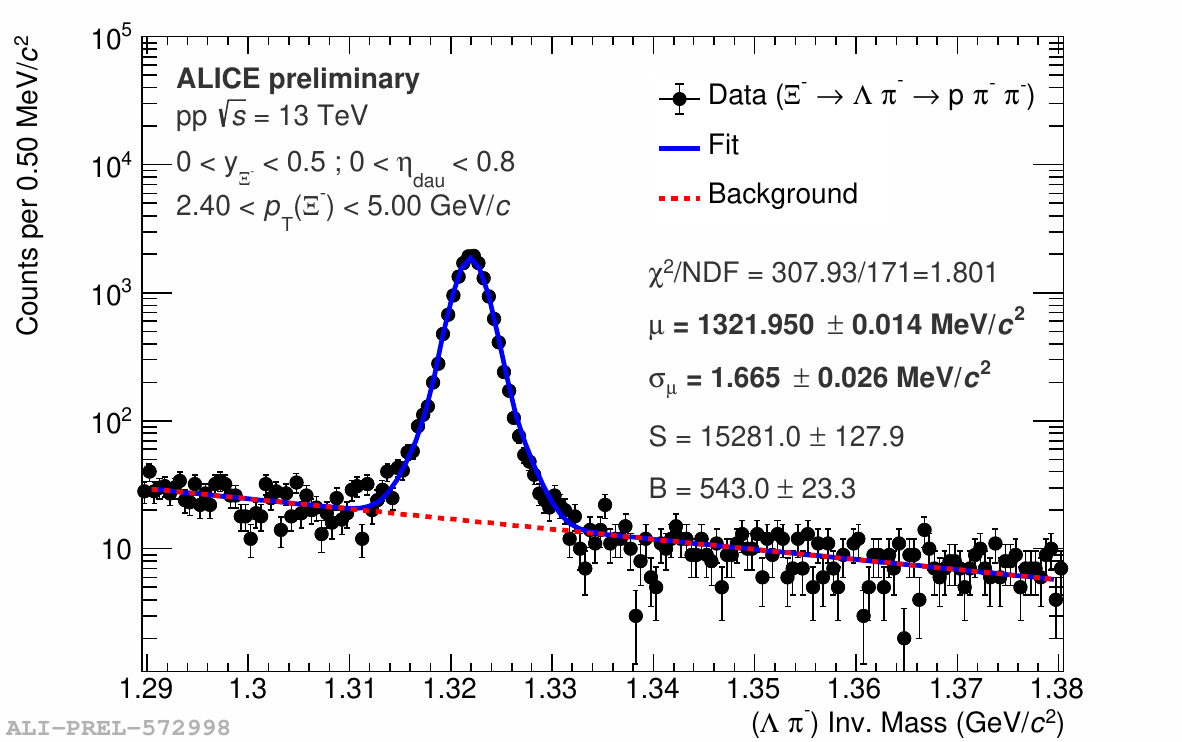}
	\label{fig:XiMinus_TripleGaussian}
}
\subfigure[]{
	\includegraphics[width=7.75cm,clip]{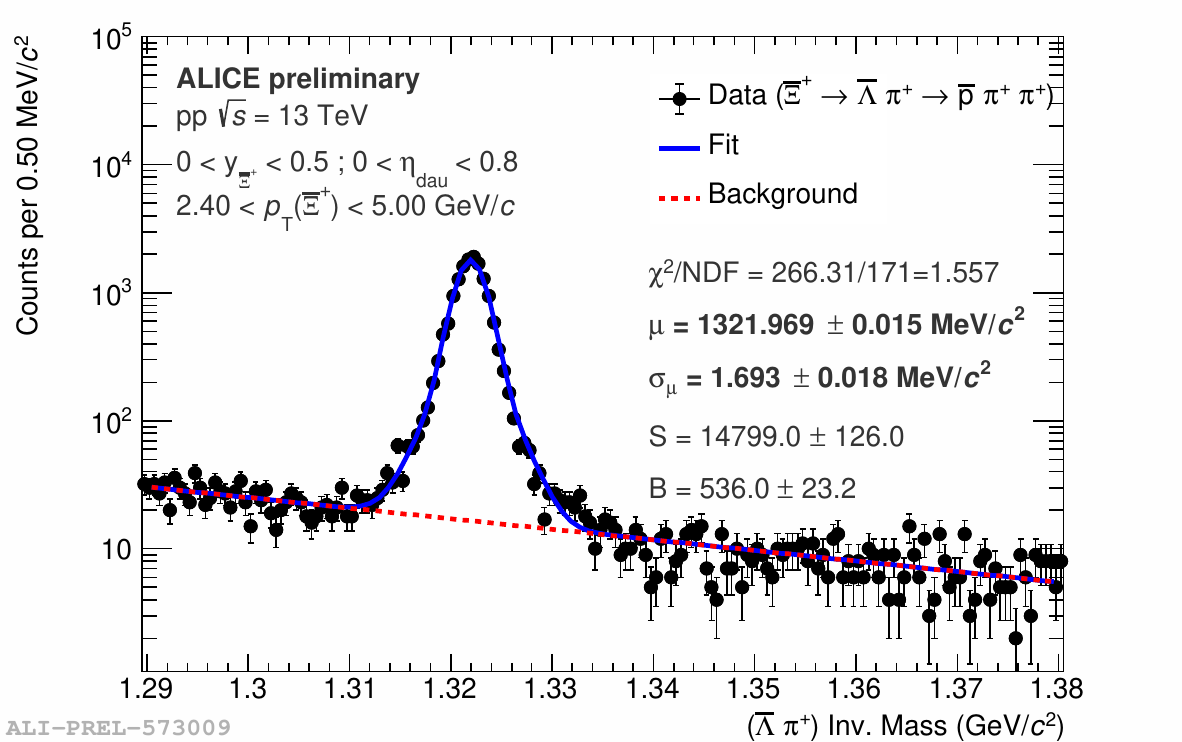}
	\label{fig:XiPlus_TripleGaussian}
} 
\subfigure[]{
	\includegraphics[width=7.75cm,clip]{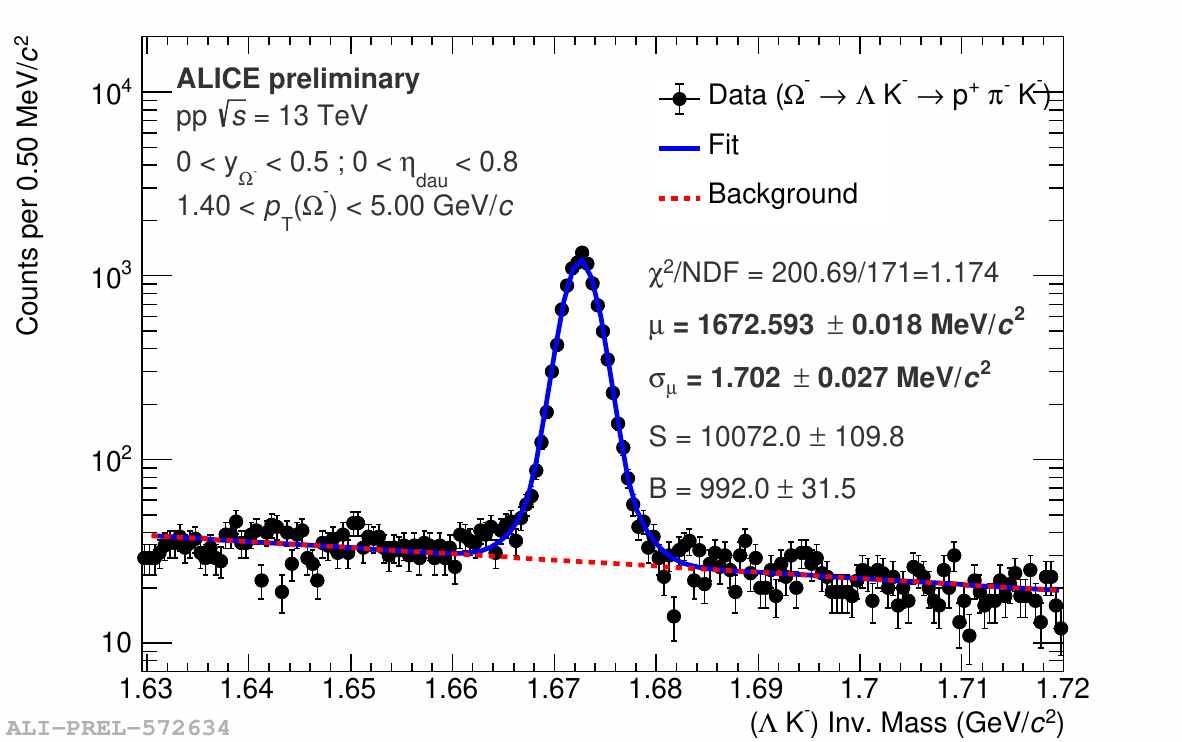}
	\label{fig:OmegaMinus_TripleGaussian}
} 
\subfigure[]{
	\includegraphics[width=7.75cm,clip]{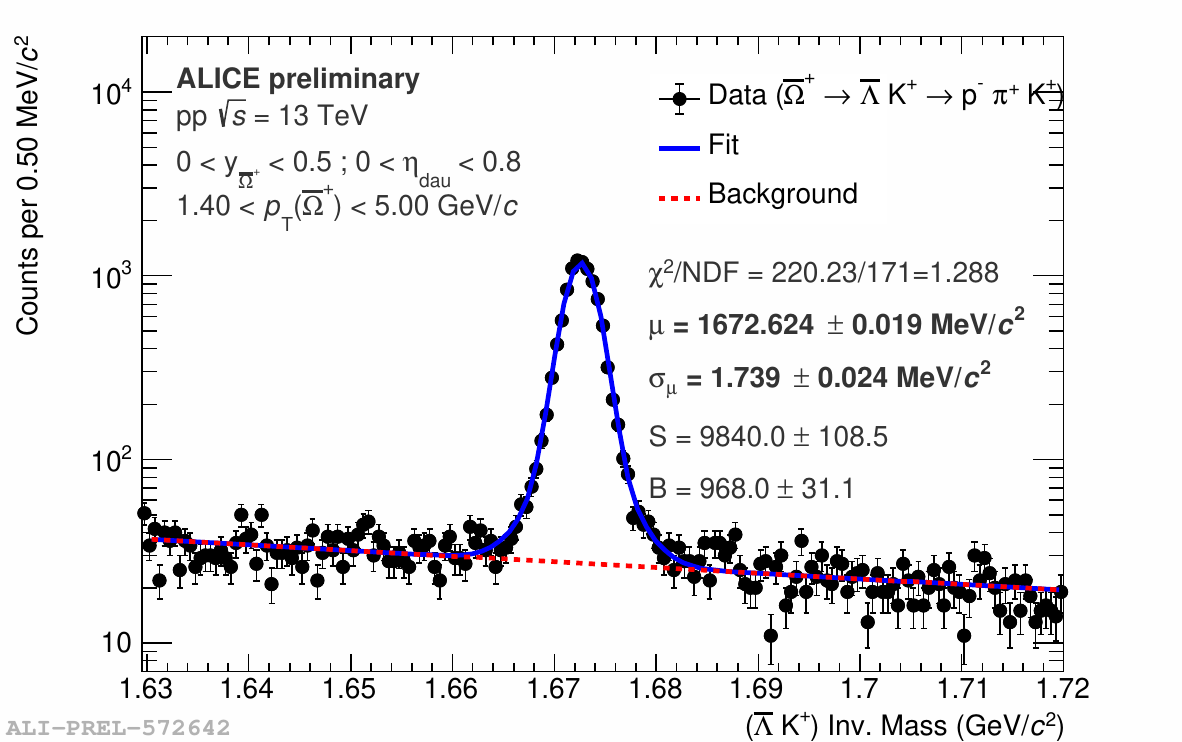}
	\label{fig:OmegaPlus_TripleGaussian}
} 
\caption{Invariant mass distributions of the $\Xi^{-}$ (\ref{fig:XiMinus_TripleGaussian}), $\overline{\Xi}^{+}$ (\ref{fig:XiPlus_TripleGaussian}), $\Omega^{-}$ (\ref{fig:OmegaMinus_TripleGaussian}) and $\overline{\Omega}^{+}$ (\ref{fig:OmegaPlus_TripleGaussian}). The peak is modelled by a triple Gaussian function, and the background by an exponential function. The measured mass and mass resolution, with their associated statistical uncertainties, are displayed in bold font.}\label{fig:InvMass}
\end{figure}

Fig.~\ref{fig:InvMass} shows the invariant mass distributions of the $\Xi^{-}$, $\overline{\Xi}^{+}$, $\Omega^{-}$, $\overline{\Omega}^{+}$ in pp collisions at $\sqrt{s} = 13$~TeV. One can see that the mass peak sits on top of a small background: 15 281 $\pm 128 \ \Xi^{-}$ (14 799 $\pm 126 \ \overline{\Xi}^{+}$) and 10 072 $\pm 110 \ \Omega^{-}$ (9 840 $\pm 109 \ \overline{\Omega}^{+}$) baryons are reconstructed, with purities reaching 96\% and 91\% respectively. The stability of the mass fits has been studied as a function of time, space, momentum, opening angles and event multiplicity. In order to control the momentum scaling,  opening angle biases, and the residual distortions in the TPC, additional selections have been implemented. For instance, for the latter case, it has been observed that the mass measurement is more stable on the positive $z$ side of the detector, hence the analysis only focuses on this region. Although, after all additional selections, only a fraction of the initial candidate sample is exploited, the measurements are still able to rely on a sample of multi-strange baryons much larger than in the previous measurements.

In order to correct for any residual bias due to the data processing, the analysis or the fit procedure, the measured masses are corrected for the mass offsets observed in Monte Carlo (MC) simulation with respect to the injected mass. The statistical error in MC is quoted as a systematic uncertainty.

\section{Study of the systematic uncertainties}

The main sources of systematic uncertainties originate from the candidate selections, the detector calibration, the finite precision on the magnetic field map and the limited knowledge on the material distribution. There are other contributors to the systematic uncertainties --- such as the mass extraction procedure, the pile-up contribution, the precision on the decay daughter masses, and the correction on the mass offset in simulation --- but these do not exceed 20~keV$/c^{2}$.

Throughout this measurement, the K$_{\rm S}^{0}$ meson and the $\Lambda$ hyperons have been used as benchmarks to ensure there is no remaining systematic bias. The final values are:
\begin{align*}
    M(\text{K}_{\rm S}^{0}) &= 497.604 \pm \text{(stat.)} 0.035 \pm \text{(syst.)} 0.254 \ \text{MeV}/c^2 ,\\
    \\
    M(\Lambda) &= 1115.775 \pm \text{(stat.)} 0.006 \pm \text{(syst.)} 0.065 \ \text{MeV}/c^2,\\
    M(\overline{\Lambda}) &= 1115.775 \pm \text{(stat.)}  0.006 \pm \text{(syst.)}  0.064 \ \text{MeV}/c^2.
\end{align*}
with a relative mass difference of
\begin{align*}
    2 \times \frac{M(\overline{\Lambda}) - M(\Lambda)}{M(\overline{\Lambda}) + M(\Lambda)} 
    &= \left[ 0.02 \pm \text{(tot.)} 2.37 \right] \times 10^{-5} .
\end{align*}
Their tabulated masses being at $M_{\rm PDG}(\text{K}_{\rm S}^{0}) = 497.611 \pm 0.013 \ \text{MeV}/c^2$ and $M_{\rm PDG}(\Lambda) = 1115.683 \pm 0.006 \ \text{MeV}/c^2$, the measured masses of K$_{\rm S}^{0}$ and $\Lambda$ agree with the PDG mass values within $2\sigma$. The same observation can be made about the relative mass difference between $\Lambda$ and $\overline{\Lambda}$\footnote{The relative mass difference between $\Lambda$ and $\overline{\Lambda}$ quoted in the PDG sits at $\left[-0.1 \pm \text{(tot.)}1.1\right]\times 10^{-5}$.}, suggesting the absence of any remaining systematic effect. 
%

\section{Results}


The final values of the $\Xi^{\pm}$ and $\Omega^{\pm}$ masses are:
\begin{align*}
    M(\Xi^{-}) &= 1321.975 \pm \text{(stat.)} 0.026 \pm \text{(syst.)} 0.078 \ \text{MeV/}c^2 ,\\
    M(\overline{\Xi}^{+}) &= 1321.964 \pm \text{(stat.)} 0.024 \pm \text{(syst.)} 0.083 \ \text{MeV/}c^2 ,\\
    M(\Omega^{-}) &= 1672.511 \pm \text{(stat.)} 0.033 \pm \text{(syst.)} 0.102 \ \text{MeV/}c^2 ,\\
    M(\overline{\Omega}^{+}) &= 1672.555 \pm \text{(stat.)} 0.034 \pm \text{(syst.)} 0.102 \ \text{MeV/}c^2 .
\end{align*}

The final relative mass difference between particle and anti-particle are:
\begin{align*}
    2 \times \frac{M(\overline{\Xi}^{+}) - M(\Xi^{-})}{M(\overline{\Xi}^{+}) + M(\Xi^{-})} &= \left[ -1.45 \pm \text{(tot.)} 6.25 \right] \times 10^{-5} ,\\
    2 \times \frac{M(\overline{\Omega}^{+}) - M(\Omega^{-})}{M(\overline{\Omega}^{+}) + M(\Omega^{-})} &= \left[ 3.28 \pm \text{(tot.)} 4.47 \right ] \times 10^{-5},
\end{align*}where the total uncertainty is calculated by summing the statistical and systematic ones in quadrature.

\section{Discussion and conclusion}

Based on a final sample of approximately 30~000~$\left( \Xi^{-} + \Xi^{+} \right)$ and 20~000~$\left( \Omega^{-} + \Omega^{+} \right)$ collected by ALICE, mass and mass difference measurements have been performed. The present results rely on a sample of strange baryons that is much larger than those in previous measurements and thus are not dominated by the statistical uncertainties anymore. By comparison with the previous measurements in Tab~\ref{tab:PDGcharac}, one can observe that the precision on the mass values has improved by 15\% in the case of $\Xi$ and by 10 fold for $\Omega$. Concerning the mass difference between particle and anti-particle, the uncertainty has reduced by 40\% and almost 2 fold, for $\Xi$ and $\Omega$ respectively. 

As a consequence of the mass difference values of multi-strange baryons being still compatible with 0, our results strengthen the single measurements quoted in the PDG, and further constraint the validity of the CPT symmetry in the multi-strange baryon sector. Furthermore, the gain in precision --- up to a factor 10 --- on the mass values will improve the input to lQCD for the physical scale determination. This is the case, for example, of the $g_{\mu}-2$ prediction of the BMW Collaboration~\cite{borsanyiLeadingHadronicContribution2021}, for which all uncertainties from the physical input would become negligible using our more precise measurements as anchor points. The shift of the multi-strange baryon masses --- the updated $\Xi$ masses being 2.5~$\sigma$ larger than the PDG values, while the $\Omega$ masses still agree with the tabulated values --- will also affect lQCD calculations. One example is related to the hadron mass spectroscopy prediction \cite{fodorLightHadronMasses2012}, which still needs to be consistent with the measured spectrum using our mass values as input. 


\bibliographystyle{JHEP}
\bibliography{Exported_Items.bib}



\end{document}